\documentclass[12pt,a4paper]{article}
\usepackage{graphicx}
\usepackage{times}
\title{\bf The Infrared Properties of Massive Stars in the Magellanic Clouds}
\author{A.Z. Bonanos$^1$ , D.J. Lennon, D.L. Massa, M. Sewilo,
  F. Koehlinger, N. Panagia, \\
  J.Th. van Loon, C.J. Evans, L.J. Smith, M. Meixner, K. Gordon and the SAGE teams
\vspace{0.8cm}\\
\normalsize $^1$
 National Observatory of Athens, IAA \\ \normalsize I. Metaxa \& Vas. Pavlou Street,
 Palaia Penteli GR-15236, Greece \\\normalsize Email: {\tt bonanos@astro.noa.gr}}
\date{\mbox{}}
\begin{document}
\maketitle
\pagestyle{empty}
%
%
\def\bull{\vrule height .9ex width .8ex depth -.1ex}
\makeatletter
\def\ps@plain{\let\@mkboth\gobbletwo
\def\@oddhead{}\def\@oddfoot{\hfil\tiny\bull\quad
``The multi-wavelength view of hot, massive stars''; 39$^{\rm th}$ Li\`ege Int.\ Astroph.\ Coll., 12-16 July 2010 \quad\bull}%
\def\@evenhead{}\let\@evenfoot\@oddfoot}
\makeatother
%
%
\def\beginrefer{\section*{References}%
\begin{quotation}\mbox{}\par}
\def\refer#1\par{{\setlength{\parindent}{-\leftmargin}\indent#1\par}}
\def\endrefer{\end{quotation}}
%
%
{\noindent\small{\bf Abstract:} We present results of our study of the
  infrared properties of massive stars in the Large and Small
  Magellanic Clouds, which are based on the Spitzer SAGE surveys of
  these galaxies. We have compiled catalogs of spectroscopically
  confirmed massive stars in each galaxy, as well as photometric
  catalogs for a subset of these stars that have infrared counterparts
  in the SAGE database, with uniform photometry from 0.3 to 24 $\mu$m
  in the UBVIJHKs+IRAC+MIPS24 bands. These catalogs enable a
  comparative study of infrared excesses of OB stars, classical Be
  stars, yellow and red supergiants, Wolf-Rayet stars, Luminous Blue
  Variables and supergiant B[e] stars, as a function of metallicity,
  and provide the first roadmaps for interpreting luminous, massive,
  resolved stellar populations in nearby galaxies at infrared
  wavelengths.}
%
%

\section{Introduction}

The {\em Spitzer Space Telescope} Legacy Surveys SAGE (``Surveying the
Agents of a Galaxy's Evolution'', (Meixner et al. 2006) and SAGE-SMC
(Gordon et al. 2010) have for the first time made possible a
comparative study of the infrared properties of massive stars at a
range of metallicities, by imaging both the Large and Small Magellanic
Clouds (LMC and SMC). In Bonanos et al. (2009, Paper I) and Bonanos et
al. (2010, Paper II), we presented infrared properties of massive
stars in the LMC and SMC, which we summarize below. The motivation was
threefold: (a) to use the infrared excesses of massive stars to probe
their winds, circumstellar gas and dust, (b) to provide a template for
studies of other, more distant, galaxies, and (c) to investigate the
dependence of the infrared properties on metallicity. Papers I and II
were the first major compilations of accurate spectral types and
multi-band photometry from 0.3$-$24 $\mu$m for massive stars in any
galaxy, increasing by an order of magnitude the number of massive
stars for which mid-infrared photometry was available.

Infrared excess in hot massive stars is primarily due to free-free
emission from their ionized, line driven, stellar winds. Panagia \&
Felli (1975) and Wright \& Barlow (1975) first computed the free-free
emission from ionized envelopes of hot massive stars, as a function of
the mass-loss rate ($\dot{M}$) and the terminal velocity of the wind
($v_{\infty}$). The properties of massive stars, and in particular
their stellar winds (which affect their evolution) are expected to
depend on metallicity ($Z$). For example, Mokiem et al. (2007) found
empirically that mass-loss rates scale as $\dot{M} \sim Z^{0.83\pm
  0.16}$, in good agreement with theoretical predictions (Vink et
al. 2001). The expectation, therefore, is that the infrared excesses
of OB stars in the SMC should be lower than in the LMC, given that
$\dot{M}$ is lower in the SMC. Furthermore, there is strong evidence
that the fraction of classical Be stars among B-type stars is higher
at lower metallicity (Martayan et al. 2007). Grebel et al. (1992) were
the first to find evidence for this, by showing that the cluster
NGC\,330 in the SMC has the largest fraction of Be stars of any known
cluster in the Galaxy, LMC or SMC. More recent spectroscopic surveys
(Martayan et al. 2010) have reinforced this result. We were also
interested in quantifying the global dependence of the Be star
fraction on metallicity.  The incidence of Be/X-ray binaries is also
much higher in the SMC than in the LMC (Liu et al. 2005), while the
incidence of Wolf-Rayet (WR) stars is much lower; therefore, a
comparison of infrared excesses for these objects was also of
interest.

\section{Spectral type and Photometric Catalogs}

We compiled catalogs of massive stars with known spectral types in
both the LMC and SMC from the literature. We then cross-matched the
stars in the SAGE and SAGE-SMC databases, after incorporating optical
and near-infrared photometry from recent surveys of the Magellanic
Clouds. The resulting photometric catalogs were used to study the
infrared properties of the stars. The LMC spectral type catalog
contains 1750 massive stars. A subset of 1268 of these are included in
the photometric catalog, for which uniform photometry from $0.3-24$
$\mu$m in the $UBVIJHK_{s}$+IRAC+MIPS24 bands is presented in Paper
I. The SMC spectral type catalog contains 5324 massive stars; 3654 of
these are included in the photometric catalog, for which uniform
photometry from $0.3-24$ $\mu$m is presented in Paper II. All catalogs
are available electronically.

\section{Infrared properties of Massive Stars}

Below we summarize some of our results on the following classes of
massive stars:

\subsection {O/Oe and early-B/Be stars}

We clearly detect infrared excesses from free-free emission despite
not having dereddened the stars, both in the LMC and SMC. In Figure 1,
we plot $J_{IRSF}$ vs.  $J_{IRSF}-[3.6]$, $J_{IRSF}-[5.8]$ and
$J_{IRSF}-[8.0]$ colors for the 1967 early-B stars from our SMC
catalog, respectively, denoting their luminosity classes, binarity and
emission line classification properties by different symbols. We
compare the observed colors with colors of plane-parallel non-LTE
TLUSTY stellar atmosphere models (Lanz \& Hubeny 2003, 2007) of
appropriate metallicity and effective temperatures. For reference,
reddening vectors and TLUSTY models reddened by $E(B-V)=0.2$ mag are
also shown.  At longer wavelengths, the excess is larger because the
flux due to free--free emission for optically thin winds remains
essentially constant with wavelength. Fewer stars are detected at
longer wavelengths because of the decreasing sensitivity of {\it
  Spitzer} and the overall decline of their SEDs. We find that the
majority of early-B supergiants in the SMC exhibit lower infrared
excesses, when compared to their counterparts in the LMC, due to their
lower mass-loss rates, although certain exceptions exist and deserve
further study.

The CMDs allow us to study the frequency of Oe and Be stars, given the
low foreground and internal reddening for the Magellanic Clouds. Our
SMC catalog contains 4 Oe stars among 208 O stars, of which one is
bluer than the rest. There are 16 additional stars with
$J_{IRSF}-[3.6]>0.5$~mag and $J_{IRSF}<15$~mag (including all
luminosity classes), whose spectra appear normal (although the
H$\alpha$ spectral region in most cases was not observed). We refer to
these as ``photometric Oe'' stars and attribute their infrared
excesses to free-free emission from a short-lived, possibly recurrent
circumstellar region, whose H$\alpha$ emission line was not detected
during the spectroscopic observations either because the gas had
dispersed or because the region was optically thick to H$\alpha$
radiation or the observation spectral range just did not extend to
H$\alpha$. Given the expectation of lower $\dot{M}$ at SMC
metallicity, we argue that such a region is more likely to be a
transient disk rather than a wind. Assuming these are all Oe stars, we
find a $10\pm2\%$ fraction of Oe stars among the O stars in the
SMC. The error in the fraction is dominated by small number
statistics. In contrast, there are 4 Oe and 14 ``photometric Oe''
stars (with $J_{IRSF}-[3.6]>0.5$~mag and $J_{IRSF}<14.5$~mag) out of
354 O stars in the LMC (despite the higher $\dot{M}$ at LMC
metallicity), which yields a $5\pm1\%$ fraction of Oe stars among O
stars in the LMC.

Turning to the early-B stars, the most striking feature in Figure~1 is
a distinct sequence displaced by $\sim0.8$~mag to the red. A large
fraction of the stars falling on this redder sequence have Be star
classifications, although not all Be stars reside there. Given that
the circumstellar gas disks responsible for the emission in Be stars
are known to completely vanish and reappear between spectra taken even
1 year apart (see review by Porter \& Rivinius 2003, and references
therein), the double sequence reported here provides further evidence
for the transient nature of the Be phenomenon. A bimodal distribution
at the $L-$band was previously suggested by the study of Dougherty et
al. (1994), which included a sample of 144 Galactic Be stars. Our
larger Be sample, which is essentially unaffected by reddening, and
the inclusion of all early-B stars, clearly confirms the bimodal
distribution. It is due to the much larger number of Be stars
classified in the SMC, in comparison to the LMC, as well as the higher
fraction of Be stars among early-B stars in the SMC, which is
$19\pm1\%$ vs.\ $4\pm1\%$ in the LMC when considering only the
spectroscopically confirmed Be stars (cf. $\sim17\%$ for $<10$~Myr
B0--5 stars; Wisniewski et al. 2006). Excluding the targeted sample of
Martayan et al. (2007a, 2007b) does not significantly bias the
statistics, since the fraction only decreases to $15\pm1\%$. We
caution that incompleteness in our catalogs could also affect the
determined fractions, if our sample turns out not to be representative
of the whole population of OB stars.

\begin{figure}[h]
\centering
\includegraphics[width=8cm, angle=270]{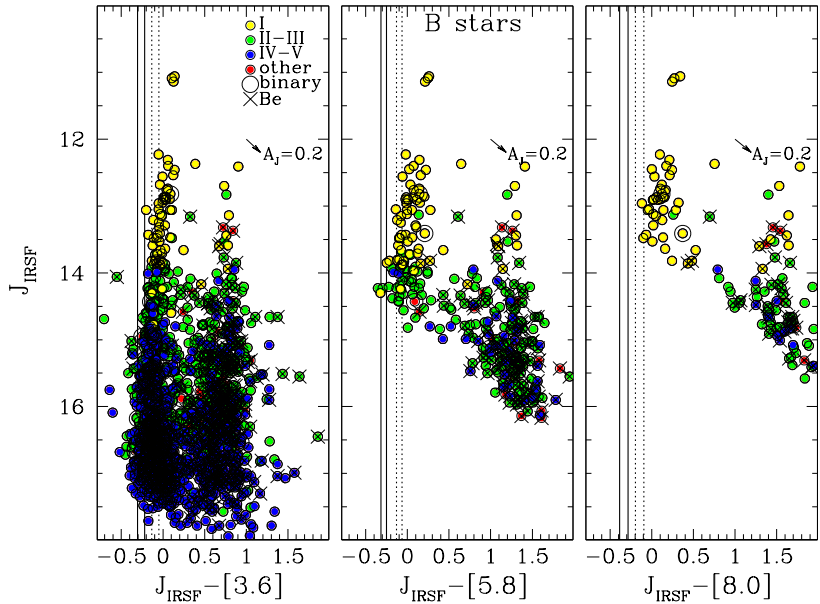}
\caption {Infrared excesses ($J_{IRSF}$ vs. $J_{IRSF}-[3.6]$,
  $J_{IRSF}-[5.8]$ and $J_{IRSF}-[8.0]$) for 1967 early-B stars in the
  SMC. Supergiants are shown in yellow, giants in green, main-sequence
  stars in blue, stars with uncertain classifications (``other'') in
  red, binaries with a large circle and Oe stars with an $\times$. The
  solid lines correspond to 30kK and 50kK TLUSTY models with $\log g =
  4.0$. A reddening vector for $E(B-V)=0.2$ mag is shown, as well as
  reddened TLUSTY models by this same amount (dotted lines). The more
  luminous stars exhibit larger infrared excesses, which increase with
  $\lambda$.}
\end{figure}

We proceed to define ``photometric Be'' stars as early-B type stars
with an intrinsic color $J_{IRSF}-[3.6]>0.5$~mag, given that a
circumstellar disk or envelope is required to explain such large
excesses. Including these ``photometric Be'' stars and using the same
color and magnitude cuts as for the ``photometric Oe'' stars above,
yields fractions of Be stars among early-B stars of $27\pm2\%$ for the
SMC and $16\pm2\%$ for the LMC (cf. 32\% from young SMC clusters;
Wisniewski et al. 2006). We compare our results with the fractions
determined by Maeder et al. (1999) from young clusters, i.e.\ 39\% for
the SMC and 23\% for the LMC, finding ours to be lower, although the
sample selections were very different.

These preliminary statistics (available for the first time for Oe
stars) indicate that both Oe and Be stars are twice as common in the
SMC than in the LMC. We emphasize the importance of including the
``photometric Be'' stars, which significantly increase the frequencies
of Oe/O and Be/early-B stars determined and are crucial when comparing
such stars in different galaxies. This novel method of confirming Oe
and Be star candidates from their infrared colors or a combination of
their optical and infrared colors, as recently suggested by Ita et
al. (2010) is complementary to the detailed spectroscopic analyses by
e.g.\ Negueruela et al. (2004) on individual Oe stars to understand
their nature, although it is limited to galaxies with low internal
reddening.  We finally note that the spectral types of Oe stars in the
SMC (O7.5Ve, O7Ve, O4-7Ve and O9-B0III-Ve) and the LMC (O9Ve (Fe II),
O7:Ve, O8-9IIIne, O3e) are earlier than those of known Galactic Oe
stars, which are all found in the O9-B0 range (Negueruela et
al. 2004).

Finally, we note that the brightest Be stars in the SMC
($J_{IRSF}\sim13.2$~mag) are brighter than the brightest Be stars in the
LMC ($J_{IRSF}\sim13.4$~mag), i.e.\ there is a 0.7 mag difference in
absolute magnitude, given the 0.5 mag difference in the distance
moduli.

\subsubsection{Supergiant B[e] stars}

The sgB[e] stars are the most conspicuous group of stars in all
infrared CMDs and TCDs: they are among the brightest and most reddened
stars in both the LMC and SMC (Buchanan et al. 2006). In the LMC, 12
stars have been classified as sgB[e] stars (including LH 85--10,
although it is not among the 11 stars listed in Zickgraf 2006). The 11
that were included in our catalog (S~22, S~134, R~126, R~66, R~82,
S~12, LH~85-10, S~35, S~59, S~137, S~93) were all matched in the SAGE
database. The SEDs of all the sgB[e] stars (except LH~85-10, which
seems to be misclassified) are all very similar, with slowly
decreasing flux in the optical, an inflexion point in the
near-infrared and a ``bump'' starting at 2~$\mu$m and peaking near
5~$\mu$m. This peak corresponds to hot dust at $\sim$600~K. The slight
change in the slopes of the SEDs between 8 and 24~$\mu$m from star to
star suggests different contributions from cool dust (150~K).

In the SMC photometric catalog, we have detected 7 luminous sources
with colors typical of sgB[e] stars, i.e.\ $M_{3.6}<-8$,
$[3.6]-[4.5]>0.7$, $J-[3.6]>2$~mag. Five of these are previously known
sgB[e] stars (with R50; B2-3[e] being the brightest in all IRAC and
MIPS bands), while R4 (AzV\,16) is classified as an LBV with a sgB[e]
spectral type. In addition to these, we find that 2dFS1804
(AFA3kF0/B[e]) has a very similar SED (and therefore infrared colors)
to the known sgB[e] 2dFS2837 (AFA5kF0/B[e]).  Evans et al. (2004) also
remarked on the similarity of their spectra. We therefore confirm the
supergiant nature of 2dFS1804. The similarity of the SEDs of these
sgB[e] stars, despite the various optical spectral classifications,
implies that all are the same class of object. The cooler, composite
spectral types indicate a lower mass and perhaps a transitional stage
to or from the sgB[e] phenomenon. The only difference we find between
the sgB[e] stars in the SMC vs. the LMC is that on average they are
$\sim$1-2 mag fainter (in absolute terms).

\subsubsection{Luminous Blue Variables}

There are 6 confirmed LBVs (see review by Humphreys \& Davidson 1994)
in the LMC: S~Dor, BAT99-83 or R127, R~71, R~110, BAT99-45, and
R~85. The LBVs are not only among the most luminous sources at
3.6~$\mu$m, with [3.6]--[4.5] colors similar to AGB stars and
intermediate between RSG and sgB[e] stars, but also at 8.0~$\mu$m and
24~$\mu$m. All 3 known LBVs in the SMC: R4 (AzV\,16, B0[e]LBV), R40
(AzV\,415, A2Ia: LBV) and HD\,5980 (WN6h;LBV binary), were detected at
infrared wavelengths. We find their SEDs to differ, given their very
different spectral types. Moreover, we find evidence for variability,
which can be confirmed from existing light curves in the All Sky
Automated Survey (ASAS, Pojmanski 2002), as pointed out by Szczygiel
et al. (2010), who studied the variability of the massive stars
presented in Paper~I in the LMC. The various SED shapes and spectral
types observed depend on the time since the last outburst event and
the amount of dust formed.

%
%
\section*{Acknowledgements}
A.Z.B. acknowledges support from the Riccardo Giacconi Fellowship
award of the Space Telescope Science Institute and from the European
Commission Framework Program Seven under a Marie Curie International
Reintegration Grant.
%
%
\footnotesize

\beginrefer

\refer {Bonanos}, A.~Z., {Massa}, D.~L., {Sewilo}, M., {et~al.} 2009,  \textit{AJ}, 138
, 1003

\refer {Bonanos}, A.~Z., Lennon, D.J., Koehlinger, F. {et~al.} 2010,  \textit{AJ}, 140,
 416

\refer {Buchanan}, C.~L., {Kastner}, J.~H., {Forrest}, W.~J., {et~al.} 2006, \textit{AJ}, 132,
  1890

\refer {Dougherty}, S.~M., {Waters}, L.~B.~F.~M., {Burki}, G., {et~al.} 1994, \textit{A\&A},
  290, 609

\refer {Evans}, C.~J., {Lennon}, D.~J., {Trundle}, C., {et~al.} 2004,
  \textit{ApJ}, 607, 451

\refer {Foellmi}, C., {Koenigsberger}, G., {Georgiev}, L., {et~al.} 2008, \textit{RevMexAA}, 44, 3

\refer {Gordon}, K.~D., {Meixner}, M., {Blum}, R., {et~al.} 2010,
\textit{AJ}, in preparation

\refer {Grebel}, E.~K., {Richtler}, T., \& {de Boer}, K.~S. 1992, \textit{A\&A}, 254, L5

\refer {Humphreys}, R.~M. \& {Davidson}, K.  1994, \textit{PASP}, 106, 1025

\refer {Ita}, Y., {Matsuura}, M., {Ishihara}, D., {et~al.} 2010a,
\textit{A\&A}, 514, 2

\refer {Lanz}, T. \& {Hubeny}, I. 2003, \textit{ApJS}, 146, 417

\refer ---. 2007, \textit{ApJS}, 169, 83

\refer  {Liu}, Q.~Z., {van Paradijs}, J., \& {van den Heuvel}, E.~P.~J. 2005,
\textit{A\&A},  442, 1135

\refer {Maeder}, A., {Grebel}, E.~K., \& {Mermilliod}, J. 1999, \textit{A\&A}, 346, 459

\refer {Martayan}, C., {Floquet}, M., {Hubert}, A.~M., {et~al.} 2007a, \textit{A\&A}, 472, 577

\refer {Martayan}, C., {Fr{\'e}mat}, Y., {Hubert}, A., {et~al.} 2007b, \textit{A\&A}, 462, 683

\refer {Martayan}, C., {Baade}, D., \& {Fabregat}, J. 2010, \textit{A\&A}, 509, A11

\refer {Meixner}, M., {Gordon}, K.~D., {Indebetouw}, R., {et~al.} 2006, \textit{AJ}, 132, 2268

\refer {Mokiem}, M.~R., {de Koter}, A., {Vink}, J.~S., {et~al.} 2007,
\textit{A\&A}, 473, 603

\refer {Negueruela}, I., {Steele}, I.~A., \& {Bernabeu}, G. 2004,
\textit{Astronomische Nachrichten}, 325, 749

\refer {Panagia}, N. \& {Felli}, M. 1975, \textit{A\&A}, 39, 1

\refer {Pojmanski}, G. 2002, \textit{Acta Astronomica}, 52, 397

\refer {Porter}, J.~M. \& {Rivinius}, T. 2003, \textit{PASP}, 115, 1153

\refer {Szczygiel}, D.~M., {Stanek}, K.~Z., {Bonanos}, A.~Z., {et~al.}
2010, \textit{AJ}, 140, 14

\refer {Vink}, J.~S., {de Koter}, A., \& {Lamers}, H.~J.~G.~L.~M. 2001, \textit{A\&A}, 369, 574

\refer {Wisniewski}, J.~P. \& {Bjorkman}, K.~S. 2006, \textit{ApJ}, 652, 458

\refer {Wright}, A.~E. \& {Barlow}, M.~J. 1975, \textit{MNRAS}, 170, 41

\refer {Zickgraf}, F.-J. 2006, in Astronomical Society of the Pacific
Conference Series, Vol. 355, Stars with the B[e] Phenomenon,
ed. M.~{Kraus} \& A.~S.  {Miroshnichenko}, 135--+

\endrefer           
\end{document}